\documentclass[fleqn,usenatbib]{mnras}

\usepackage{newtxtext,newtxmath}
\usepackage[T1]{fontenc}

\DeclareRobustCommand{\VAN}[3]{#2}
\let\VANthebibliography\thebibliography
\def\thebibliography{\DeclareRobustCommand{\VAN}[3]{##3}\VANthebibliography}

\usepackage{graphicx}	% Including figure files
\usepackage{amsmath}	% Advanced maths commands
\usepackage{amssymb}	% Extra maths symbols

\title[Revisiting sticking property of pebbles]
{Revisiting sticking property of submillimetre-sized aggregates}

\author[Sota Arakawa]{
Sota Arakawa$^{1,2}$\thanks{E-mail: sota.arakawa@nao.ac.jp}
\\
$^{1}$Department of Earth and Planetary Sciences, Tokyo Institute of Technology, Meguro, Tokyo, 152-8551, Japan.\\
$^{2}$Division of Science, National Astronomical Observatory of Japan, Mitaka, Tokyo, 181-8588, Japan.
}

\date{Accepted 2020 June 16. Received 2020 June 4; in original form 2020 May 5}

\pubyear{2020}

\begin{document}
\label{firstpage}
\pagerange{\pageref{firstpage}--\pageref{lastpage}}
\maketitle

\begin{abstract}
Understanding the physical properties of dust aggregates is of great importance in planetary science.
In this study, we revisited the sticking property of submillimetre-sized aggregates.
We revealed that the ``effective surface energy'' model used in previous studies underestimates the critical pulling force needed to separate two sticking aggregates.
We also derived a new and simple model of the critical pulling force based on the canonical theory of two contacting spheres.
Our findings indicate that we do not need to consider the ``effective surface energy'' of dust aggregates when discussing the physical properties of loose agglomerates of submillimetre-sized aggregates.
\end{abstract}


%%%%%%%%%%%%%%%%%%%%%%%%%%%%%%%%%%%%%%%%%%%%%%%%%%

%%%%%%%%%%%%%%%%% BODY OF PAPER %%%%%%%%%%%%%%%%%%

\section{Introduction}

The formation of planetesimals is initiated by the collisional growth of small dust grains and aggregates in protoplanetary disks.
The collision velocities are small enough for initially $\mu$m-sized dust grains to be able to stick together following a collision, and thereby, larger dust aggregates are formed.
However, as the aggregates grow, the collision velocities increase, which makes it more difficult for more dust particles to stick to the aggregates under the conditions in the solar nebula \citep[e.g.,][]{Blum+2008}.
Therefore, revealing the transition between sticking and bouncing regimes for colliding dust aggregates is of great importance in the context of planet formation.

The suborbital particle and aggregation experiment was carried on the REXUS 12 suborbital rocket \citep[see][]{Brisset+2013}.
The sticking properties of submillimetre-sized aggregates were measured by \citet{Brisset+2016}.
They observed a growth of dust aggregates and the formation and fragmentation of clusters of up to a few millimetres in size.
The transition from bouncing to sticking collisions happened at collision velocities of around $10\ {\rm cm}\ {\rm s}^{-1}$ for dust aggregates composed of monodisperse monomer grains of $\sim 1\ {\mu}{\rm m}$ in size \citep{Brisset+2016}.

The experimental results of \citet{Brisset+2016} were interpreted using the dust aggregate model proposed by \citet{Weidling+2012}.
\citet{Weidling+2012} developed a microgravity experiment and carried out experiments of free collisions between dust aggregates with diameters of $0.5$--$2\ {\rm mm}$.
They also attempted to develop a dust aggregate model to explain their experimental results by introducing an ``effective surface energy'' of millimetre-sized dust aggregates (see Section \ref{sec.ESE}).

The effective surface energy model \citep{Weidling+2012} has been widely used in recent studies of loose agglomerations of dust aggregates \citep[e.g.,][]{Gundlach+2012,Blum+2017,Hu+2019}.
This is because recent numerical simulations and laboratory experiments suggest that planetesimals formed in the gaseous solar nebula have a pebble-pile structure \citep[e.g.,][]{Blum2018}.
In the solar nebula, millimetre- and centimetre-sized dust aggregates called ``pebbles'' may be formed via collisional growth and compaction \citep[e.g.,][]{Guettler+2010}.
These pebbles can clump together through streaming instabilities, and form gravitationally bound ``pebble clouds'' \citep[e.g.,][]{Carrera+2015,Yang+2017}.
Pebbles in a cloud will undergo dissipative mutual collisions \citep[e.g.,][]{WahlbergJansson+2014}, and finally turn into pebble-pile planetesimals, i.e., loose agglomerates of pebbles.

However, the physical background of the effective surface energy model is unclear.
In this study, we revisit the sticking properties of submillimetre-sized aggregates studied by \citet{Brisset+2016}.
We reveal that the effective surface energy model underestimates the critical pulling force needed to separate two sticking aggregates (see Section \ref{sec.results}).
We also derive a simple new model of the critical pulling force based on the canonical theory of two contacting spheres \citep{Johnson+1971} in Section \ref{sec.MMC}.

\section{Critical pulling force needed to separate two sticking monomers}

First, we review the critical pulling force needed to separate two spherical monomer grains, which forms the basis of the theory of the critical pulling force needed to separate two sticking aggregates.
Based on the JKR theory \citep*{Johnson+1971} for elastic, deformable spheres, the critical pulling force needed to separate two sticking spheres, $F_{\rm crit}$, is given by
\begin{equation}
F_{\rm crit} = \frac{3}{2} \pi \gamma r,
\label{eq.Fcrit}
\end{equation}
where $\gamma$ is the material surface energy and $r$ is the monomer radius \citep[see also][]{Wada+2007}.
\footnote{
Note that \citet{Bradley1932} and \citet{Derjaguin1934} showed that the critical pulling force between two rigid spheres is equal to $F_{\rm crit} = 2 \pi \gamma r$, instead of ${( 3/2 )} \pi \gamma r$ \citep[see][]{Greenwood2007}.
}
In this study, we assume that all monomer particles have the same radius of $r$.
It has been experimentally confirmed that the above equation of $F_{\rm crit}$ is applicable to micron-sized ${\rm Si}{\rm O}_{2}$ spherical particles \citep[e.g.,][]{Heim+1999}.

\citet{Johnson+1971} stressed that the critical pulling force needed to separate two sticking spheres, $F_{\rm crit}$, is independent of the Young's modulus of the monomers, $E$.
Therefore, $F_{\rm crit}$ is independent of the contact radius, $a_{\rm c}$.
The contact radius of two sticking monomer grains is given by
\begin{equation}
a_{\rm c} = {\left[ \frac{9 \pi \gamma {\left( 1 - \nu^{2} \right)}}{2 E r} \right]}^{1/3} r,
\label{eq.ac}
\end{equation}
where $\nu$ is Poisson's ratio \citep[e.g.,][]{Johnson+1971,Wada+2007}.
The material properties of ${\rm Si}{\rm O}_{2}$ grains are listed in \citet{Dominik+1997}: $\gamma = 25\ {\rm mJ}\ {\rm m}^{-2}$, $E = 54\ {\rm GPa}$, and $\nu = 0.17$.

\section{Critical pulling force needed to separate two sticking aggregates}
\label{sec.agg}

In Section \ref{sec.agg}, we introduce two models for the critical pulling force needed to separate two sticking aggregates, namely, the ``effective surface energy'' model \citep{Weidling+2012} and the monomer--monomer contact model (Section \ref{sec.MMC}, this study).
The critical pulling force for separating two sticking submillimetre-sized aggregates, $F_{\rm c}$, was experimentally obtained by \citet{Brisset+2016}.
In their experiments, the statistical threshold velocity between the sticking and bouncing regimes of aggregates was measured.
Then, the critical pulling force $F_{\rm c}$ was evaluated based on the threshold velocity between the sticking and bouncing regimes \citep[see][for details]{Brisset+2016}.

\subsection{Effective surface energy model}
\label{sec.ESE}

Here, we briefly summarize the ``effective surface energy'' model.
\citet{Weidling+2012} introduced the effective surface energy, $\gamma_{\rm eff}$, as a combination of the material surface energy, $\gamma$, the filling factor of aggregates, $\phi$, and the Hertz factor (i.e., the ratio between the contact surface and the cross section of two monomer grains), ${a_{\rm c}}^{2} / r^{2}$.
The effective surface energy is given by
\begin{equation}
\gamma_{\rm eff} = \gamma \phi \frac{{a_{\rm c}}^{2}}{r^{2}}.
\label{eq.geff}
\end{equation}

In the framework of the effective surface energy model, the critical pulling force is given by 
\begin{equation}
F_{\rm c}^{\rm ESE} = N \cdot \frac{3}{2} \pi \gamma_{\rm eff} r_{\rm agg},
\label{eq.ESE}
\end{equation}
where $N$ is the number of inter-aggregate connections per aggregate and $r_{\rm agg}$ is the aggregate radius.
When a dust cluster consists of two aggregates, the number of inter-aggregate connections per aggregate is $N = 1$.
The upper limit of $N$ is $12$, which is the maximum coordination number for the packing of monodisperse dust aggregates.
\citet{Brisset+2016} obtained $N \simeq 2.5$ from snapshot images of dust clusters during their experiment.

The filling factor of dust aggregates used in \citet{Brisset+2016} is $\phi = 0.37$, in which case the critical pulling force is given by
\begin{equation}
F_{\rm c}^{\rm ESE} = N \cdot 1.5 \times 10^{-9}\ {\left( \frac{\gamma}{25\ {\rm mJ}\ {\rm m}^{-2}} \right)}^{5/3} {\left( \frac{r}{1\ {\mu}{\rm m}} \right)}^{- 2/3} {\left( \frac{r_{\rm agg}}{0.1\ {\rm mm}} \right)}\ {\rm N}.
%1.49565...
\end{equation}
However, the experimental results suggest that $F_{\rm c} \simeq 10^{-7}\ {\rm N}$ for submillimetre-sized aggregates (see Figure \ref{fig2}).
As the maximum coordination number for monodispersed dust aggregates in a dust cluster is $N = 12$, we conclude that the effective surface energy model \citep{Weidling+2012} cannot explain the critical pulling force of submillimetre-sized aggregates.
%\footnote{
%{\bf
%We also note that \citet{Brisset+2016} estimated the material surface energy, $\gamma$, based on the effective surface energy, $\gamma_{\rm eff}$, obtained from experiments; however, their result contains errors (see Appendix \ref{app}).
%}
%}

\subsection{Monomer--monomer contact model}
\label{sec.MMC}

\begin{figure}
\centering
\includegraphics[width = 0.6\columnwidth]{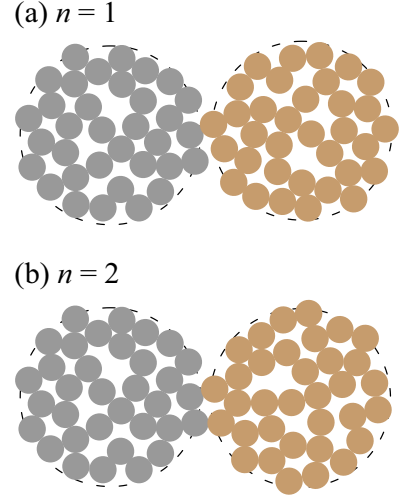}
\caption{
Schematic illustrations of two sticking aggregates.
(a) Two sticking aggregates with inter-aggregate contacts $n = 1$.
(b) Two sticking aggregates with inter-aggregate contacts $n = 2$.
We assume that $n \sim 1$ for the contacts between two sticking aggregates.
}
\label{fig1}
\end{figure}

Here, we propose a new simple model of $F_{\rm c}$ for two sticking aggregates.
Figure \ref{fig1} shows a schematic illustration of two sticking aggregates.
When there is no external pressure, the number of inter-aggregate contacts of monomers, $n$, is $n \sim 1$.
Then, the critical pulling force of two sticking aggregates is given by
\begin{align}
F_{\rm c}^{\rm MMC} &= n \cdot \frac{3}{2} \pi \gamma r, \nonumber \\
                    &= n \cdot 1.2 \times 10^{-7}\ {\left( \frac{\gamma}{25\ {\rm mJ}\ {\rm m}^{-2}} \right)} {\left( \frac{r}{1\ \mu{\rm m}} \right)}\ {\rm N},
\label{eq.MMC}
\end{align}
which is equal to the $F_{\rm crit}$ of two sticking monomers when we assume $n = 1$.
We found that the monomer--monomer contact model can explain the experimental results of \citet{Brisset+2016}.

\section{Results}
\label{sec.results}

\begin{figure}
\centering
\includegraphics[width = \columnwidth]{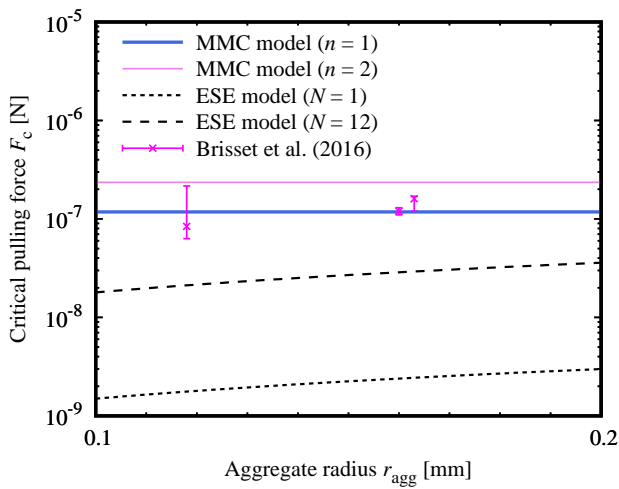}
\caption{
The critical pulling force needed to separate two sticking submillimetre-sized aggregates is $F_{\rm c}$.
Magenta crosses with a vertical error bar denote the experimental results of \citet{Brisset+2016}.
Blue and violet solid lines are the predictions of the monomer--monomer contact model, $F_{\rm c}^{\rm MMC}$.
We assumed the number of inter-aggregate contacts of monomers to be $n = 1$ (blue) or $n = 2$ (violet).
The black dashed lines represent the predictions based on the effective surface energy model, $F_{\rm c}^{\rm ESE}$, for $N = 1$ and $N = 12$ \citep[e.g.,][]{Weidling+2012,Brisset+2016}.
}
\label{fig2}
\end{figure}

Figure \ref{fig2} shows the experimental results of $F_{\rm c}$ for various aggregate radii $r_{\rm agg}$.
The monomers used in \citet{Brisset+2016} are $1\ \mu{\rm m}$-sized ${\rm Si}{\rm O}_{2}$ grains, and we set $r = 1\ \mu{\rm m}$ in our calculations of $F_{\rm c}^{\rm MMC}$ and $F_{\rm c}^{\rm ESE}$.

We found that the effective surface energy model underestimates the critical pulling force by order(s) of magnitude, even if we set $N = 12$, which is the theoretical upper limit of the number of inter-aggregate connections per aggregate.
In contrast, the predictions of $F_{\rm c}$ of the monomer--monomer contact model are consistent with the experimental results of \citet{Brisset+2016}, especially for the case of $n = 1$.
We can imagine that separations of sticking monomers are not necessarily simultaneous, and that the lower limit of $F_{\rm c}$ needed to separate two sticking aggregates is equal to $F_{\rm crit}$, that is, the critical pulling force needed to separate two sticking monomers.

We acknowledge that there is a large uncertainty in the estimate of the material surface energy of ${\rm Si}{\rm O}_{2}$ grains.
\citet{Kimura+2015} and \citet{Steinpilz+2019} claimed that the surface energy of hydrophilic amorphous silica grains depends on the surface chemistry (i.e., the water content) of the grains; the surface energy of dry samples ($\sim 200\ {\rm mJ}\ {\rm m}^{-2}$) is one order of magnitude higher than that of wet samples ($\sim 25\ {\rm mJ}\ {\rm m}^{-2}$).
If the material surface energy is $\gamma \sim 200\ {\rm mJ}\ {\rm m}^{-2}$, the effective surface energy model may reproduce the experimental results of \citet{Brisset+2016}.
However, we note that spherical monomer grains used in their experiments are amorphous silica samples manufactured by Micromod Partikeltechnologie GmbH \citep[see][]{Kothe+2013}, and several layers of water molecules might be present on the grain surface when they did not heat samples to remove surface water before experiments \citep{Steinpilz+2019}.
Therefore, we assumed that the material surface energy of ${\rm Si}{\rm O}_{2}$ grains used in their experiments is $\gamma \sim 25\ {\rm mJ}\ {\rm m}^{-2}$.

\section{Discussion}
\label{sec.discussion}

It is known that the adhesion force between elastic solids is affected by their surface roughness, and the roughness usually reduces the adhesion force \citep[e.g.,][]{Fuller+1975}.
The significant reduction of the adhesion force of contacting dust aggregates in comparison with smooth spheres is also experimentally confirmed \citep[e.g.,][]{Weidling+2012,Brisset+2016,Demirci+2019}.
These facts might be the motivation of \citet{Weidling+2012} to introduce the effective surface energy model.

\citet{Fuller+1975} revealed that the total adhesion force is obtained by applying the JKR theory to each individual asperity, as we assumed in the monomer--monomer contact model.
The number of asperities is identical to $n$ of Equation (\ref{eq.MMC}).
Their model predicts that the adhesion force depends on the ratio between $\sigma$ and $\delta$, where $\sigma$ is the deviation of asperity heights and $\delta$ is the elastic displacement.
\citet{Fuller+1975} showed that the total adhesion force increases with decreasing the ratio, $\sigma / \delta$ \citep[see Figure 6 of][]{Fuller+1975}.
For the case of two sticking dust aggregates, the deviation of asperity heights may be approximately given by the monomer radius, $\sigma \sim r$.
On the other hand, the elastic displacement of dust aggregates should depend on the effective Young's modulus of dust aggregates and the external force applied to contacting aggregates \citep[e.g.,][]{Weidling+2012}.
Then $\delta$ increases with increasing applied force.
In addition, the increase of the applied force also enlarges the contact radius of two sticking aggregates.
Therefore, the total adhesion force increases with increasing applied force because the number of asperities, $n$, increases.

We note that \citet{Blum+2014} measured the tensile strength of loose agglomerates of pebbles, and they found that the tensile strength depends on the applied compression.
We will discuss the tensile strength of loose agglomerates of pebbles in future study.

\section{Conclusion}

Understanding the sticking property of submillimetre-sized aggregates is of great importance in planetary science.
Using the REXUS 12 suborbital rocket, \citet{Brisset+2016} performed a microgravity experiment of collisional sticking and fragmentation of submillimetre-sized aggregates of $1\ \mu{\rm m}$-sized ${\rm Si}{\rm O}_{2}$ grains.

\citet{Brisset+2016} reported the critical pulling force needed to separate two sticking aggregates, $F_{\rm c}$, and we compared two theoretical models, the effective surface energy model \citep[e.g.,][]{Weidling+2012} and the monomer--monomer contact model (Section \ref{sec.MMC}, this study), with the experimental results.

We found that the effective surface energy model underestimates the critical pulling force by order(s) of magnitude, while the monomer--monomer contact model is consistent with the experimental results (see Figure \ref{fig2}).
Therefore, we do not need to consider the ``effective surface energy'' of pebbles (millimetre- and centimetre-sized dust aggregates) when evaluating the physical properties of pebble-pile planetesimals.

\section*{Acknowledgements}

The author sincerely thanks Taishi Nakamoto, Shigeru Ida, Satoshi Okuzumi, Kenji Ohta, and Hidenori Genda for insightful comments and careful reading of the draft.
The author also thanks the anonymous reviewer for prompt and constructive comments.
This work was supported by JSPS KAKENHI Grants (JP17J06861, JP20J00598) and by the Publications Committee of NAOJ.

\section*{Data availability}

The data underlying this article will be shared on reasonable request to the corresponding author.

%%%%%%%%%%%%%%%%%%%%%%%%%%%%%%%%%%%%%%%%%%%%%%%%%%



\bibliographystyle{mnras}
\bibliography{references} % if your bibtex file is called example.bib

\bsp	% typesetting comment
\label{lastpage}
\end{document}